# Optimization of Design Parameters for SPPC Longitudinal Dynamics

**L.H. Zhang,**[a, b, c] **J.Y. Tang,**[a, b, c,1] **Y. Hong,**[a, b, c] **Y.K. Chen**[a, c] **and L.J. Wang**[a, b, c]

[a] *Institute of High Energy Physics, Chinese Academy of Sciences,*
  *Yuquan Road 19B, Shijingshan District, Beijing, 100049, China*

[b] *University of Chinese Academy of Sciences,*
  *Yuquan Road 19A, Shijingshan District, Beijing, 100049, China*

[c] *Spallation Neutron Source Science Center,*
  *No.1, Zhongziyuan Road, Dalang, Dongguan, 523803, China*
  *E-mail*: `tangjy@ihep.ac.cn`

ABSTRACT: As the second stage of the CEPC-SPPC project, SPPC (Super Proton-Proton Collider) aims at exploring new physics beyond the Standard Model. The key design goal for the SPPC accelerator complex is to reach 75 TeV in center of mass energy with a circumference of 100 km for the collider and an injector chain of four accelerators in cascade to support the collider. As an important part of the SPPC conceptual study, the longitudinal beam dynamics was studied systematically, which includes the dynamics in the collider and its complex injector chain. First, the bunch filling scheme of the SPPC complex was designed on the basis of various constraints such as the technical challenges of the kicker magnets, limited extraction energy per injection and so on. Next, the study on the longitudinal dynamics in the collider focused on the RF scheme to meet the requirements for luminosity and mitigate relevant instabilities. A higher harmonic RF system (800 MHz) together with the basic RF system (400 MHz) to form a dual-harmonic RF system was employed to mitigate collective instabilities and increase the luminosity by producing shorter bunches. In addition, the longitudinal matchings between the bunches in the different accelerator stages were studied, with special attention to the space charge effects and the beam loading effect in the two lower energy rings (p-RCS and MSS), which result in an optimization of the RF schemes. A set of self-consistent beam and RF parameters for the SPPC complex was obtained. The collider and the three proton synchrotrons of the injector chain have unprecedented features, thus this study demonstrates how a future proton-proton collider complex looks like.

---

[1] Corresponding author.

# Contents

# 1. Introduction

Following the discovery of the Higgs boson at the Large Hadron Collider (LHC) in 2012, which opens a brand-new door for the unknown physics, next-generation large colliders are being proposed and studied by the international high-energy community to explore the Higgs boson in depth and probe new physics beyond the Standard Model. In China, a two-stage project including two colliders of an unprecedented scale – CEPC and SPPC (Circular Electron-Positron Collider & Super Proton-Proton Collider) was initiated [1-2], with CEPC (Phase I) focusing on the Higgs physics and SPPC (Phase II) being an energy frontier collider or a discovery machine which is far beyond the parameter reach of the LHC. The two colliders share the same tunnel of 100 km in circumference. While CEPC-SPPC is comparable to the CERN-based project FCC, SPPC competes with FCC-hh [3].

The key design goal for the SPPC is to achieve 75 TeV in center-of-mass energy with 12-T superconducting magnets of iron-based high-temperature technology and to obtain the nominal luminosity of $1\times10^{35}$ cm$^{-2}$s$^{-1}$. Some key parameters of the SPPC baseline design are presented in table 1. Figure 1 depicts the layout of the SPPC accelerator complex and functions of the SPPC main rings (double-ring structure for two opposite-direction proton beams). There are eight identical arcs and eight long straight sections represented respectively by LSS1_coll for a complicated collimator system, LSS2_rf for RF stations, LSS3_pp and LSS7_pp for two proton-proton collisions, LSS4_AA for possible ion-ion collision, LSS5_ext for extraction, LSS6_inj for injection and LSS8_ep for possible electron-proton collision.

To pre-accelerate the beam to the injection energy of 2.1 TeV of the SPPC main ring with the required beam properties such as bunch current, bunch structure, emittance and beam fill period, a four-stage injector chain is designed: a superconducting proton linac (p-Linac) of 1.2 GeV with a repetition rate of 50 Hz, a rapid cycling synchrotron (p-RCS) boosting the energy to 10 GeV with a repetition rate of 25 Hz, a medium-stage synchrotron (MSS) to 180 GeV with a lower repetition rate of 0.5 Hz and a final super synchrotron (SS) to 2.1 TeV. Table 2 shows the main parameters for the SPPC injector chain [2]. Since only fractions of the operation time for the different stages are needed to supply beams to the SPPC, they could operate with longer duty cycles to provide higher-power beams for other research applications when they do not prepare beams for the SPPC.

As the longitudinal beam dynamics in all the five accelerator stages are strongly related, it was studied systematically and presented here. First, the bunch filling scheme in the collider rings to meet the target luminosity is shown in section 2. Then, section 3 presents the longitudinal dynamics parameters for the collider rings, with focus on the RF scheme to meet the requirements for luminosity and to mitigate relevant instabilities. Besides, the longitudinal dynamics parameter design on the injector chain is described in section 4. Section 5 presents the conclusions and discussion.

**Table 1.** Some key parameters of SPPC baseline design.

| Parameter | Value | Unit |
|---|---|---|
| Circumference | 100 | km |
| Injection energy | 2.1 | TeV |
| Collision energy | 37.5 | TeV |
| Number of IPs | 2 | |
| Nominal luminosity | 1.0×10³⁵ | cm⁻²s⁻¹ |
| Dipole field at collision | 12 | T |
| Bunch population | 1.5×10¹¹ | |
| Bunch spacing | 25 | ns |
| RF frequency | 400 | MHz |
| RMS bunch length at collision | 7.55 | cm |

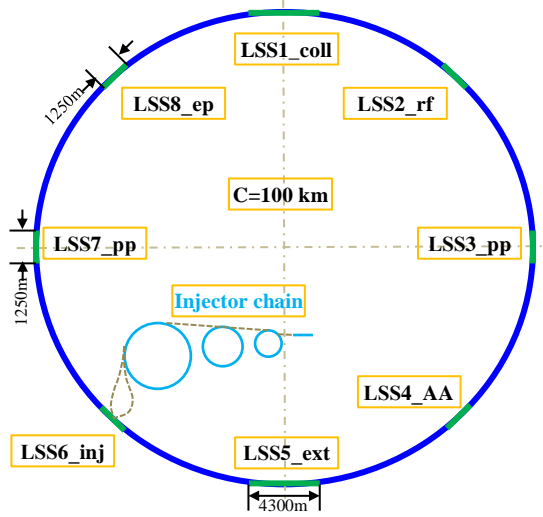

**Figure 1.** SPPC accelerator complex.

**Table 2.** Main parameters for the SPPC injector chain.

| Parameter | Unit | p-Linac | p-RCS | MSS | SS |
|---|---|---|---|---|---|
| Energy | GeV | 1.2 | 10 | 180 | 2100 |
| Average current | mA | 1.4 | 0.34 | 0.02 | - |
| Length/Circumference | km | ~0.3 | 0.97 | 3.5 | 7.2 |
| Repetition rate | Hz | 50 | 25 | 0.5 | 1/30 |
| Ramp-up time | s | - | 0.02 | 0.8 | 12 |
| Max. beam power or energy | MW/MJ | 1.6/ | 3.4/ | 3.7/ | /34 |
| Dipole field | T | - | 1.0 | 1.7 | 8.3 |
| Beam fraction for next stage | % | 50 | 6 | 13.3 | >2 |

## 2. Bunch filling scheme

The bunch filling scheme is an important part of the SPPC longitudinal dynamics, which is to accommodate a maximum number of bunches in the collider to achieve as high as possible luminosity. As the bunch filling scheme has direct relationship to the bunch spacing, the injection and extraction gaps for each stage of the accelerator complex, it was studied together with the whole accelerator chain. Furthermore, the bunch filling scheme has also a strong impact on the multi-bunch beam dynamics and beam-beam effects in the collider.

## 2.1 Requirements and constraints on bunch filling

### 2.1.1 Nominal bunch spacing and the bunch filling factor

The bunch spacing does not contribute to the final bunch filling factor of the SPPC main rings, but a shorter bunch spacing does help accumulate relatively more bunches with smaller bunch particles, which is highly beneficial to alleviate the beam-beam effects and reduce the event pileup per crossing in the detectors but with some sacrifice of the initial luminosity. On the other hand, the bunch spacing is also limited by both the parasitic collisions in the proximity of Interaction Points (IPs) and the electron cloud instability. The nominal bunch spacing of 25 ns at SPPC has been adopted in the baseline design and is used in this study, though other bunch spacing of 10 ns and 5 ns are also under study. The final bunch filling factor is required to be as close to 0.8 as possible, similar to LHC [4].

The nominal bunch spacing of 25 ns is formed in the second injector stage or p-RCS with an RF system close to 40 MHz, and a bunch splitting mechanism can be applied in MSS if needed.

### 2.1.2 Beam-beam effects

In order to avoid missing head-on collision and minimize the number of Pacman bunches, the symmetry of the bunch trains in the SPPC should be generated as far as possible [4, 5]. And to keep the long-range interaction points fixed, the injection gaps should be multiples of the basic bunch spacing of 25 ns [6].

### 2.1.3 Injection/extraction gaps

The injection and extraction systems, especially the kicker rise times, impose critical constraints on the arrangement of the bunches or bunch trains in the different stages of the SPPC complex. The time gaps between the bunch trains are essential for beam injection and extraction in all the four ring-type accelerators. Their lengths depend on the practical technical design of injection and extraction systems. Usually for higher kinetic energy, longer rise time for the kickers is needed. Taking into account the future technology development, the most demanding time gap that is for the beam extraction in the SPPC is set to 3 μs.

The energy stored in beam in SS can be as high as 34 MJ. To avoid the device damage in case of abnormal extraction from SS or injection into SPPC, the beam is intentionally extracted in multiple times from SS, for example, the energy per extraction is 5.65 MJ in six times, which is close to the FCC-hh design of 5 MJ. However, this will create more time gaps in SS and SPPC that decrease the bunching filling factor.

## 2.2 Beam pulsing structure from p-Linac to p-RCS

The p-Linac is a superconducting linear accelerator with the repetition rate of 50 Hz and RF duty factor of 3.4%. H- ions are accelerated to 1.2 GeV in the p-Linac, then stripped into protons by the stripping foil located at the injection region of the p-RCS, and finally filled into the RF buckets of the p-RCS by both transverse and longitudinal phase-space paintings. The nominal bunch population of $1.5\times10^{11}$ for the SPPC is also defined here. Figure 2 represents the evolution process of beam pulsing structure from p-Linac to p-RCS.

A macro-pulse beam with a length of 0.68 ms is first produced from the ion source at the repetition period of 20 ms, then forms into micro pulses or bunches with a period of 3.08 ns in the RFQ with the RF frequency of 325 MHz. Two types of choppers with different frequencies of 36 MHz and 0.28 MHz, respectively, in the low/medium-energy beam transfer lines (LEBT/MEBT) are needed to achieve the low-loss phase space painting in the longitudinal place in p-RCS and to create a large time gap for the extraction. Chopper 1 with a repetition rate of 36 MHz that corresponds to the RF frequency of p-RCS at the injection energy has a chopping rate of 50% or half micro bunches removed. Thus, a series of medium pulses of 27.78 ns in period are formed. Chopper 2 with a repetition rate of 0.28 MHz that corresponds to the revolution frequency of RCS at injection will further create a series of long pulses of 3.57 μs in period with a time gap of 444.48 ns or equivalent 16 medium pulses for extraction. Finally, 112 bunches out of 128 RF buckets in p-RCS are filled. The total injection turns are about 191 corresponding to the macro pulse length of 0.68 ms in p-Linac.

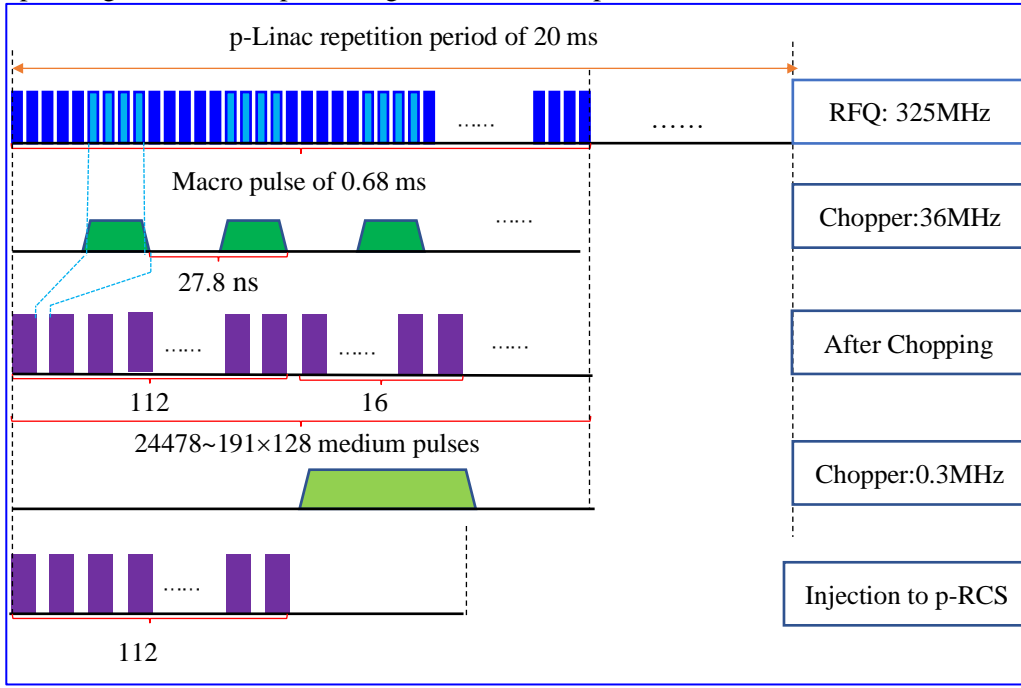

**Figure 2.** Beam pulsing structure from p-Linac to p-RCS.

### 2.3 Bunch patterns from p-RCS to SPPC

The bunch patterns from p-RCS to SPPC are illustrated in figure 3.

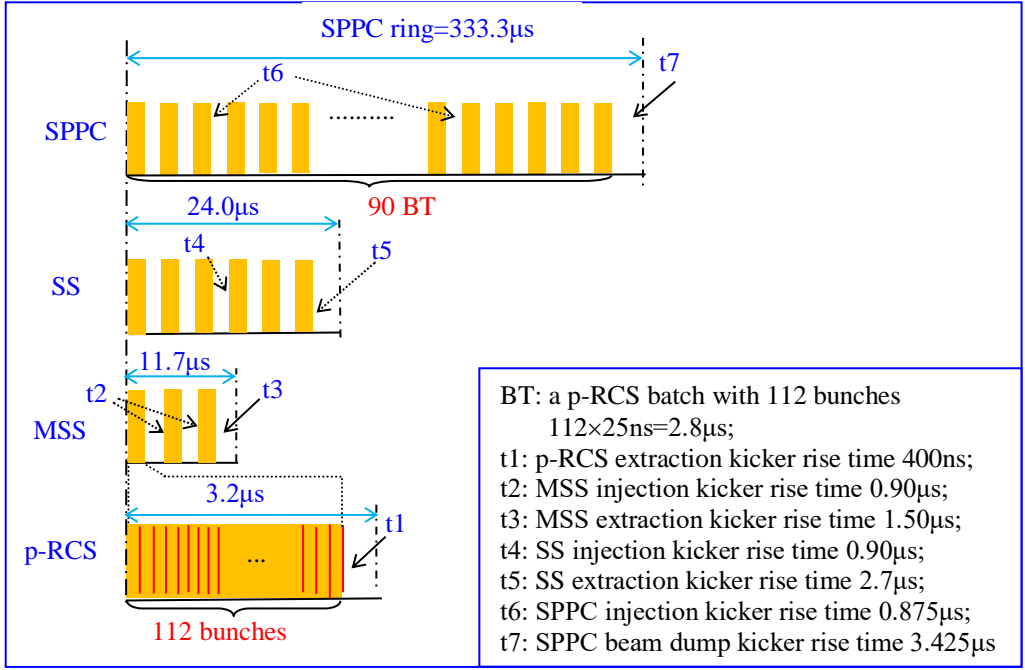

**Figure 3.** The bunch filling scheme from p-RCS to SPPC.

### 2.3.1 The p-RCS

During the acceleration process from 1.2 GeV to 10 GeV in p-RCS, the RF frequency shifts gradually from 36 MHz to 40 MHz following the change of the beam revolution frequency. At the extraction, the nominal bunch spacing of 25 ns, the length of the bunch train (abbr. as BT) of 2.8 μs including 112 bunches and the time gap (t1) of 400 ns for the extraction kickers are formed. The bunch train is extracted in a single turn and injected into MSS by the bunch-to-bucket method, which will be depicted in detail in section 4.1.

### 2.3.2 The MSS

With the kinetic beam energy approaching to the velocity of light, the RF frequency of 40 MHz in MSS keeps almost constant during acceleration. The revolution period is 11.7 μs. The MSS ring is filled with 3 injections from the p-RCS with a time gap of 0.9 μs for the injection kickers, and a time gap of 1.5 μs for extraction is left. The stored beam of 3 BTs is extracted in a single turn and injected into SS. Hence, the bunch structure in the MSS ring can be described in the following form: BT + 0.9 μs + BT + 0.9 μs + BT + 1.5 μs.

### 2.3.3 The SS

As mentioned above, due to the very high stored energy in beam up to 34 MJ for the whole SS beam at the extraction, the 6 bunch trains in the SS has to be extracted in six times to prevent damage in the case of abnormal functioning in the transfer beamline from SS to SPPC and in the SPPC injection.

A basic RF system with a constant frequency of 200 MHz is utilized in the SS, but the basic bunch spacing is the same as that of the upstream accelerators, p-RCS and MSS, which means only one of every five RF buckets is filled. The revolution period in the SS ring is 24.0 μs. The SS ring is stuffed with 2 injections from the MSS with a time gap of 0.9 μs for the injection kickers and the extraction kickers in the normal operation. A time gap of 2.7 μs for the rise time of the special extraction kickers that extract the whole beam to an external beam dump in one turn in the abortion case is reserved, and a much longer flat-top time is also needed for these kickers. Therefore, the bunch structure in the SS ring can be depicted in the following form: BT + 0.9 μs + BT + 0.9 μs + BT + 0.9 μs + BT + 0.9 μs + BT + 0.9 μs + BT + 2.7 μs.

### 2.3.4 The SPPC

The basic RF system with a constant frequency of 400 MHz is adopted for the SPPC. The revolution period is about 333.3 μs, which will require 15 SS batches or 90 BTs for each of the two rings to be transferred. The SPPC injection kicker rise time is 0.875 μs and identical for each injection, which leaves the time interval of 3.425 μs for the SPPC extraction kicker rise time. Thus, the bunch structure in the SPPC ring can be represented as follows: 89 × (BT + 0.875 μs) + BT + 3.425 μs. A total of 10080 bunches are injected into the SPPC ring, and the bunch filling factor of SPPC reaches about 75.6%, which is slightly lower than 80% at LHC but similar to FCC-hh [7] and considered acceptable. Shorter rise time of SPPC injection kickers helps boost the bunch filling factor, which will make the technical design of injection kickers more challenging. With the SS cycle of 30 s, the filling of each SPPC ring will take about 7.5 minutes.

## 3. Parameter design of the longitudinal dynamics in the collider ring

For the parameter design of the longitudinal dynamics in the SPPC collider ring, the main goal is to obtain a set of self-consistent RF and beam parameters to achieve the target bunch length for the SPPC baseline scheme. The parameters in table 1 are used as input, and the possibility of a shorter bunch length for luminosity upgrade is considered.

### 3.1 Analytical expressions of the longitudinal beam dynamics with a dual-harmonic RF system

The dual-harmonic RF system can work in different operation modes, such as the bunch-shortening mode (BSM) or the bunch-lengthening mode (BLM), depending on the phase difference between the two RF systems [8-10]. It can also work in the single-harmonic RF mode (SRF) when the RF voltage ratio of the doubled-harmonic RF system to the fundamental RF system or $k$ is set to zero. Here, we only consider the case of $k$=0.5 for both BLM and BSM under the stationary RF bucket. Figure 4 shows the RF buckets in the normalized longitudinal phase space for SRF, $k$=0.5 for both BLM and BSM.

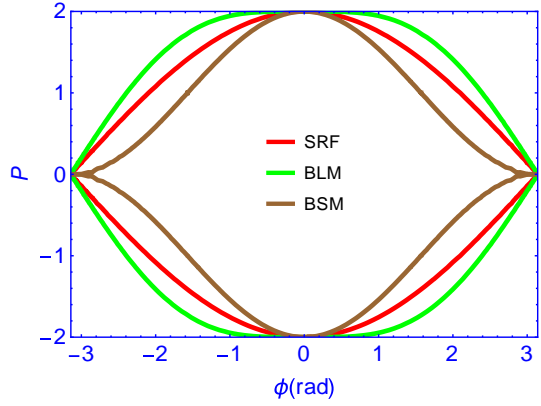

**Figure 4.** The RF buckets in the normalized longitudinal phase space for SRF, *k*=0.5 for both BLM and BSM, with $P = -(h|\eta|Q_s)/(\Delta p/p)$ the normalized momentum coordinate.

The RF bucket height can be represented by the maximum relative energy spread as:

$$\left(\frac{\Delta E}{E_0}\right)_{max} = \beta\sqrt{\frac{2eV_0}{\pi h|\eta|E_0}}, \tag{3.1}$$

where $V_0$ is the voltage of the fundamental RF system, $h$ is the harmonic number of the fundamental harmonic RF, $\eta = \alpha_p - 1/\gamma^2$ is the phase slippage factor with $\alpha_p$ the momentum compaction factor, $\gamma$ the relative energy, $\beta$ the relative velocity, and $E_0$ the energy of synchrotron particle. Eq. (3.1) can be applicable to the cases of SRF, BSM, and BLM with $k \leq 0.5$.

The area of an RF bucket that is enclosed by the separatrix in the longitudinal phase space can be expressed by [11]:

$$A_{bk} = \frac{16}{h\omega_0}\sqrt{\frac{\beta^2 E_0 eV_0}{2\pi h|\eta|}}\alpha_b, \tag{3.2}$$

where $\omega_0$ is the revolution angular frequency of the synchronous particle, the coefficient $\alpha_b = 1$ for SRF, $\alpha_b = 1.1478$ for BLM and $\alpha_b = 0.7854$ for BSM.

In practice, bunched particles only occupy a fraction of the RF bucket which is called the longitudinal emittance or bunch area, and it can be represented by the outermost single-particle emittance in the adiabatic sense. In the small-amplitude oscillation approximation, the RMS bunch length $\sigma_z$ and the momentum spread $\sigma_\delta$ can be expressed, for the SRF mode, by [11]:

$$\sigma_z = \frac{R}{2h}\left(h\varepsilon_s\sqrt{\frac{2h\omega_0^2|\eta|}{\pi\beta^2 E_0 eV_0}}\right)^{\frac{1}{2}}, \tag{3.3}$$

$$\sigma_\delta = \frac{\omega_0}{2\beta^2 E_0}\left(\frac{\varepsilon_s}{\pi}\sqrt{\frac{h\beta^2 E_0 eV_0}{2\pi\omega_0^2|\eta|}}\right)^{\frac{1}{2}}, \tag{3.4}$$

with $\varepsilon_s$ the whole bunch area. Here, we assume that the RMS bunch area $\varepsilon_{rms}$ is a quarter of the whole bunch area $\varepsilon_s$. For BSM, the RMS bunch length and the momentum spread become:

$$\sigma_z = \frac{R}{2h}\left(h\varepsilon_s\sqrt{\frac{h\omega_0^2|\eta|}{\pi\beta^2 E_0 eV_0}}\right)^{\frac{1}{2}}, \tag{3.5}$$

$$\sigma_\delta = \frac{\omega_0}{2\beta^2 E_0}\left(\frac{\varepsilon_s}{\pi}\sqrt{\frac{h\beta^2 E_0 eV_0}{\pi\omega_0^2|\eta|}}\right)^{\frac{1}{2}}, \tag{3.6}$$

with $R$ the average radius of the ring. For BLM, the RMS bunch length and the momentum spread become [12]:

$$\sigma_z = \frac{R}{2h}\left(\frac{\omega_0 h\varepsilon_s}{0.87402}\sqrt{\frac{h\pi|\eta|}{2\beta^2 E_0 eV_0}}\right)^{\frac{1}{3}}, \tag{3.7}$$

$$\sigma_\delta = \frac{\omega_0}{8\beta^2 E_0}\left(\frac{h\varepsilon_s}{0.87402}\left(\frac{2\beta^2 E_0 eV_0}{h\pi\omega_0^2|\eta|}\right)^{\frac{1}{4}}\right)^{\frac{2}{3}}, \tag{3.8}$$

Particles bounded within the RF bucket perform synchrotron oscillations around the synchronous phase $\phi_s$. Under the small-amplitude oscillation approximation, the synchrotron frequency for SRF and BSM can be evaluated by the one of the zero-amplitude particle, as:

$$f_s = f_0 \cdot Q_s = f_0 \cdot a\sqrt{\frac{heV_0|\eta|}{2\pi\beta^2 E_0}}, \tag{3.9}$$

where $Q_s$ is the synchrotron tune, $f_0$ is the revolution frequency of the synchronous particle, the coefficient $a = 1$ for SRF, whereas $a = \sqrt{2}$ for BSM [9]. On the other hand, the synchrotron frequency of the zero-amplitude particle will be zero for BLM, which will be shown in figure 11. Hence, the synchrotron frequency with the amplitude of the RMS bunch length is used and can be expressed by [10]:

$$f_s = \frac{\pi}{\sqrt{2}K}\frac{h\sigma_z}{R}\cdot f_0\sqrt{\frac{heV_0|\eta|}{2\pi\beta^2 E_0}}, \tag{3.10}$$

where $K \approx 1.8541$ is the value of the complete elliptic integral of first kind with modulus $k \approx 1/\sqrt{2}$.

The bunch occupation of a bucket can be represented by the momentum filling factor, which is usually limited to be less than 0.8 during the acceleration cycle to avoid beam loss. The momentum filling factor, which is defined by the ratio of the maximum momentum spread in the bunch to the bucket height [13, 14], can be expressed as:

$$q_p = \sqrt{\sin^2\frac{\phi_m}{2} \pm \frac{k}{2}\sin^2\phi_m}, \tag{3.11}$$

where $\phi_m$ is the maximum amplitude of the phase oscillation for the outermost particle, $k = 0$ is for SRF, '+' is for BSM and '-' for BLM, which is valid for $k \leq 0.5$. For single-harmonic RF system, under the small-amplitude approximation with $\phi_m = 2\sigma_\phi = 2\,h\sigma_z/R$, eq. (11) will be simplified as:

$$q_p \approx \frac{h\sigma_z}{R}, \tag{3.12}$$

Synchrotron radiation can normally be neglected in proton synchrotrons. However, in very high energy colliders, like the LHC and the SPPC, its effect must be considered. The energy loss per turn by synchrotron radiation for a proton with kinetic energy $E_0$ in the unit of GeV is [11]:

$$U_0(\text{keV}) = 7.783 \times 10^{-12} \times \frac{E_0^4(\text{GeV})}{\rho(m)}, \quad (3.13)$$

Here, $\rho$ is the radius of the bending magnet in the unit of meter. And the longitudinal emittance damping time due to the synchrotron radiation is:

$$\tau_E = \frac{1}{2}\frac{T_0 E_0}{U_0}, \quad (3.14)$$

with $T_0$ the revolution period of the synchronous particle.

### 3.2 Influence of bunch length on luminosity

The luminosity is expressed as [15]:

$$L = \frac{f_0 n_b N_b^2}{4\pi\varepsilon\beta^*} = \frac{I\gamma\xi}{e\beta^* r_p}, \quad (3.15)$$

Here, $I = ef_0 n_b N_b$ is the circulation beam current with $N_b$ the number of particle per bunch (bunch population) and $n_b$ the number of bunches per beam, $\beta^*$ is the beta function at the IP, $\xi = N_b r_p / 4\pi\varepsilon_n$ is the beam-beam parameter with $r_p \approx 1.53 \times 10^{-18}$ m being the classical proton radius and $\varepsilon_n$ the normalized transverse emittance. To avoid the parasitic collisions near the IP, most colliders use a crossing angle at the IP, which could reduce significantly the luminosity without countermeasures. The luminosity reduction factor due to the crossing angle can be expressed, assuming the distance of the first parasitic collision point equal to 12 times the beam envelop of that location, as [15]:

$$F_{ca} = \frac{1}{\sqrt{1 + (\frac{\sigma_z \theta_c}{2\sigma^*})^2}} = \frac{1}{\sqrt{1 + \Phi^2}}, \quad (3.16)$$

$$\Phi = \frac{\sigma_z \theta_c}{2\sigma^*} \approx 12\sqrt{\frac{\sigma_z^2}{(c\Delta t)^2} + \frac{1}{4(\beta^*/\sigma_z)^2}}, \quad (3.17)$$

Here, $\Phi$ is the Piwinski angle with $\theta_c$ the full crossing angle, $\sigma^*$ the transverse beam size at the IP, $c$ the speed of light and $\Delta t$ the bunch spacing. When the influence of bunch length and the rapid change of the beta function around the IP are taken into consideration, the luminosity will further be decreased, which is the so-called hourglass effect, by a reduction factor:

$$F_h = \frac{\beta^*}{\sqrt{\pi}\sigma_z} \exp(\frac{\beta^{*2}}{2\sigma_z^2}) K_0(\frac{\beta^{*2}}{2\sigma_z^2}), \quad (3.18)$$

where $K_0$ is the 0[th] order modified Bessel function of the second kind. Therefore, the influence of the hourglass effect and the crossing angle is illustrated in figure 5. One can see that the hourglass effect can be neglected if $\beta^*/\sigma_z \gg 1$, which is the case for the SPPC baseline scheme. A relatively large value of $\beta^*/\sigma_z$ is important to avoid significant luminosity loss caused by the crossing angle. Therefore, for the luminosity upgrade phase when we plan to decrease the $\beta^*$, it is better to shorten the RMS bunch length at the same time, which is one of the goals of the longitudinal dynamics design.

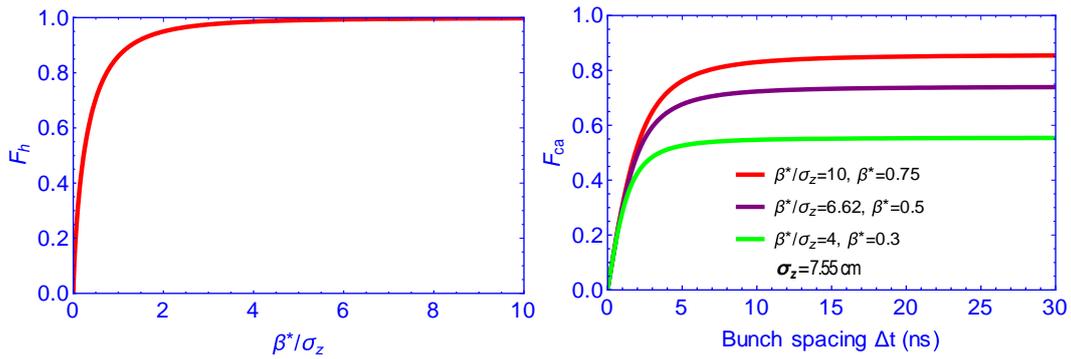

**Figure 5.** The luminosity reduction factor due to the hourglass effect (left) and the crossing angle (right).

### 3.3 Different effects in the longitudinal dynamics

In such a high-energy proton collider like the SPPC, synchrotron radiation becomes non-negligible, especially in the collision phase of the acceleration cycle. Additionally, since a relatively large bunch population of $1.5 \times 10^{11}$ protons will be utilized to achieve the nominal luminosity of $1 \times 10^{35}$ cm$^{-2}$s$^{-1}$, beam instabilities become a major concern. Therefore, the longitudinal dynamics study has to take the synchrotron radiation damping and controlled emittance blow-up, beam instabilities and mitigations into consideration. According to the phenomena observed and the experience accumulated during the SPS and LHC design and operation [8, 16, 17], the main single bunch collective effects in the SPPC should be the intra-beam scattering, the loss of Landau damping and the transverse mode coupling instability. More details are presented below.

#### 3.3.1 Intra-beam scattering

The Intra-Beam Scattering (IBS) effect is a multiple small-angle Coulomb scattering of charged particles within a bunch. It normally leads to two effects in proton or ion accelerators: redistribution of beam momenta and impact on beam quality due to transverse and longitudinal emittance growth. In this study for SPPC, the code MADX [18] was used to compute the emittance growth time due to intra-beam scatterings. And the IBS growth times at the injection and collision are shown in figure 6. In the injection phase, the longitudinal emittance between 1.5 eVs and 2.5 eVs seems to be a good choice because the IBS growth time of at least 20 h is much greater than the injection time of 840 s. However, in the collision phase, the longitudinal emittance has to be increased to at least 6 eVs due to the physics collision time as long as about 10 h and the short longitudinal emittance damping time of 1.07 h. Therefore, a controlled emittance blow-up scheme during the acceleration cycle is crucial. Thus, the IBS is well under control in the whole acceleration cycle, provided that the longitudinal emittance in the collision phase is maintained to a high value in the presence of strong radiation damping. Furthermore, the controlled emittance blow-up is also needed for suppressing other beam instabilities.

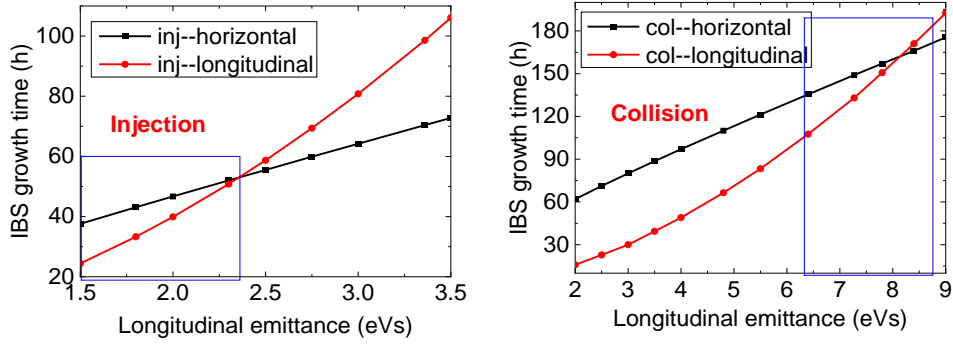

**Figure 6.** IBS growth time at SPPC injection (left) and collision (right).

### 3.3.2 Loss of Landau damping

The Landau damping in the longitudinal phase plane comes from the spread in the synchrotron frequency due to the nonlinearity of the RF voltage. This is a natural stabilizing mechanism for different longitudinal instabilities. The loss of Landau damping caused by the reactive impedance overtaking the threshold of Landau damping has been observed in LHC and SPS [8, 19] and the stability condition can be expressed by the following formula [20]:

$$\frac{|Im\,Z|}{n} \leq \frac{3\pi^2}{32}\frac{h^3 V_{rf}}{I_b}\left(\frac{L}{C}\right)^5, \qquad (3.19)$$

where $I_b = ef_0 N_b$ is the single bunch current, $V_{rf}$ is the RF voltage, $L = 4\sigma_z$ is the full bunch length, and $C$ is the circumference of the ring. Thus, the threshold on the longitudinal reactive impedance varying with the bunch length at injection and at collision is shown in figure 7. In the injection phase, the RMS bunch length is at least 8 cm in order to have the longitudinal impedance threshold greater than 0.2 Ω and to match the upstream accelerator of the injector chain in the longitudinal phase plane. Hence, a lower RF voltage should be used. However, in the collision phase, to reach the RMS bunch length of 7.55 cm that is required by the luminosity in the SPPC baseline design, an RF voltage at least 32 MV is needed and a similar limitation on the impedance threshold above 0.2 Ω is maintained.

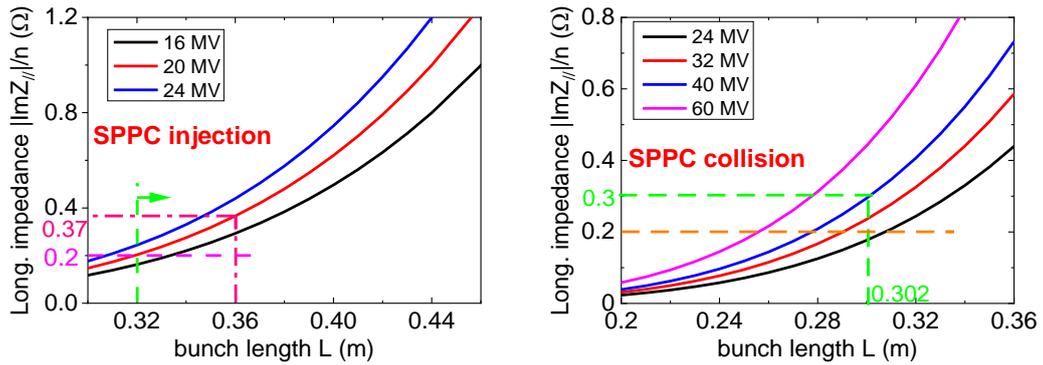

**Figure 7.** Longitudinal reactive impedance threshold at SPPC injection (left) and collision (right).

### 3.3.3 Transverse mode coupling instability (TMCI)

As the bunch intensity increases, the interaction between different head-tail modes becomes non-negligible. Since the tune shift of a mode increases with the bunch intensity, it can encounter and couple with a higher order mode, and then suddenly a strong instability occurs, which is called the Transverse Mode-Coupling Instability (TMCI) [21]. The growth of the instability can be as fast as the synchrotron oscillation period. The corresponding impedance threshold when this instability occurs can be evaluated by [20]:

$$\beta_{av} \text{Im}(Z_T)_{eff} \leq 2 \frac{E_0}{e} \frac{Q_s}{I_b} \frac{L}{R}, \tag{3.20}$$

where $\beta_{av}$ is the average beta function along the whole ring. For the SPPC, $\beta_{av}$ is 208 m at injection whereas 290 m at collision. There is a difference on $\beta_{av}$ between injection and collision due to different lattice designs for the two IP regions (LSS3_pp and LSS7_pp). Figure 8 illustrates the variation of the transverse impedance threshold with bunch length at injection and in collision. At injection, the requirement of the RMS bunch length of at least 8 cm corresponds to the transverse impedance threshold of greater than 4.5 GΩ when RF voltage is above 16 MV. However, at collision, the impedance threshold will increase to at least 25 GΩ for the RMS bunch length of 7.55 cm when the RF voltage is above 32 MV. Therefore, more stringent constraint on the transverse impedance is at injection. A larger synchrotron tune can help mitigate this instability, which is also a major concern for the parameter design of longitudinal dynamics.

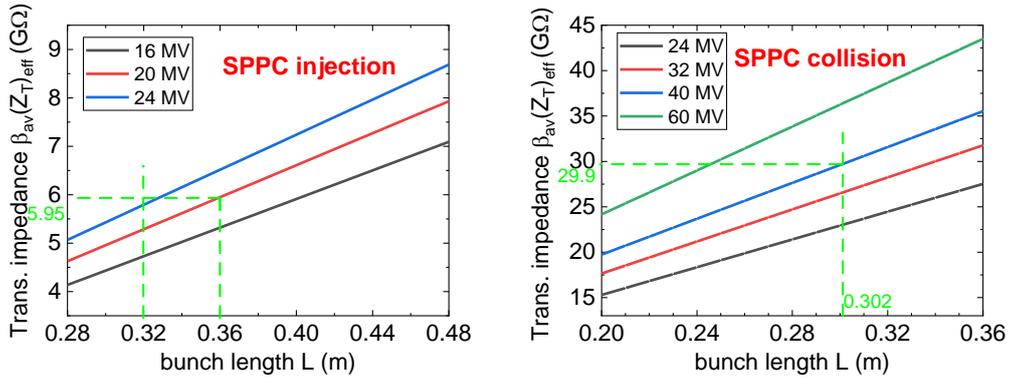

**Figure 8.** Transverse impedance threshold at the SPPC injection (left) and collision (right).

### 3.4 Basic design for the RF and beam parameters

RF parameters like RF voltage and RF frequency play key roles in the determination of longitudinal beam dynamics parameters, like the bunch length and momentum spread. The choice of RF voltage is related to the longitudinal emittance during acceleration cycle and the momentum filling factor $q_p$. And the longitudinal emittance (or bunch length) is restricted by the different single-bunch collective effects, as mentioned in section 3.3.

For a single-harmonic RF system, the relation of between momentum spread and the bunch length can be expressed, with eq. (3.3) and eq. (3.4), as:

$$\sigma_\delta = \frac{\sigma_z}{\beta c} \frac{\omega_0}{|\eta|} \sqrt{\frac{heV_0|\eta|}{2\pi\beta^2 E_0}}. \tag{3.21}$$

Using eq. (3.4) and eq. (3.21), the longitudinal emittance becomes:

$$\varepsilon_s = 4\pi\sigma_z^2 \frac{E_0}{c^2} \frac{\omega_0}{|\eta|} \sqrt{\frac{heV_0|\eta|}{2\pi\beta^2 E_0}}. \tag{3.22}$$

Thus, considering the requirements of the SPPC baseline design on the RMS bunch length of 7.55 cm and the RF frequency of 400 MHz, the momentum spread and the longitudinal emittance varying with RF voltage in SPPC collision can be illustrated in figure 9. Both of the momentum spread and the longitudinal emittance are proportional to the square root of RF voltage for a given bunch length.

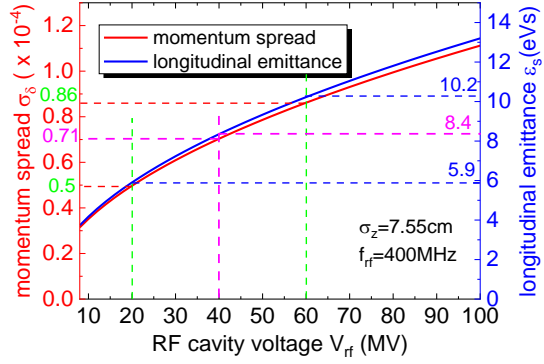

**Figure 9.** The momentum spread and the longitudinal emittance varying with RF voltage for the case of SPPC collision.

The influence of RF frequency and RF voltage on the bunch length in the SPPC collision for a given momentum spread is shown in figure 10 (left) with eq. (3.21). One can see that the bunch length is inversely proportional to the square root of the RF voltage. To achieve the designed RMS bunch length of 7.55 cm, the RF voltage required for the larger RF frequency will be lower. Therefore, the RF system at 800 MHz could be an alternative to the RF system at 400 MHz. Additionally, both the transverse and longitudinal impedance thresholds for the 800-MHz RF system can be improved due to a larger average synchrotron tune and synchrotron frequency spread within the bunch in comparison with the 400-MHz RF system. However, the RMS bunch length will have to be reduced to 5.56 cm or less from 7.55 cm in the case of 400 MHz, which happens to be advantageous to the luminosity, due to the limit of the momentum filling factor below 0.8 during physics running to avoid beam loss, as illustrated in figure 10 (right).

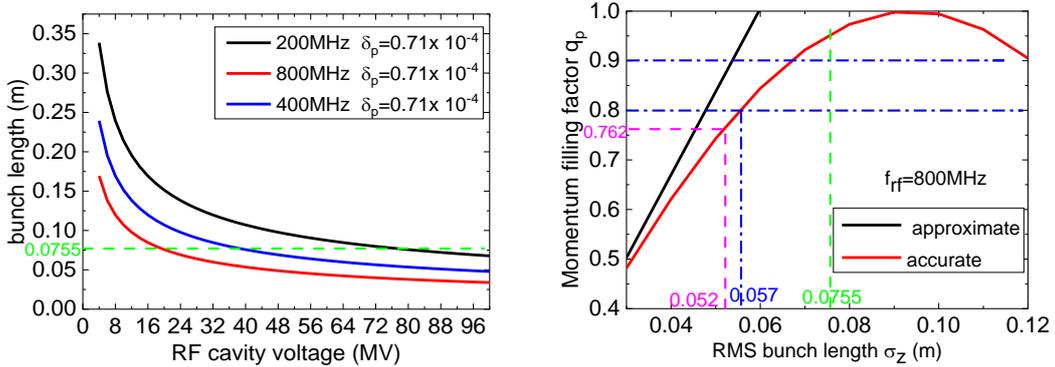

**Figure 10.** The influence of RF frequency and RF voltage on the bunch length in SPPC collision energy (left); The relation between momentum filling factor and the bunch length for the 800-MHz RF system (right).

The main acceleration system at SPPC adopts 400-MHz superconducting cavities. The RF powers delivered to beam during acceleration and for the compensation of synchrotron radiation in the begining of collision are about 8.5 MW and 1.1 MW per ring, respectively. The 400-MHz or 800-MHz single-harmonic RF system can be utilized for physics running, as described above. For an easy and perfect longitudinal phase-space matching with the SS, an additional RF system of 200 MHz in the SPPC is considered helpful during injection phase, and the required RF voltage is only 3.7 MV, as shown in table 3 (Inj-1). After the RF capture, the 400-MHz RF system is initiated, and the voltage is adiabatically increased to 20 MV, as shown in table 3 (Inj-2), and the voltage of 200-MHz RF system is gradually decreased to zero. It is worth noting that the narrow-band impedance and the resulting high-order modes (HOM) caused by 200-MHz RF cavity may add the risk of leading to the coupled bunch instability and be disadvantageous to the long-term physics running, but it can be mitigated either by tuning the cavity away from the harmful resonant frequency, or by damping this resonance mode using various ways [22-24].

Based on all the considerations above, a set of longitudinal dynamics parameters for SPPC baseline scheme with different combinations of RF frequencies at injection and collision are designed, as presented in table 3. For the scheme of 400-MHz RF system, the longitudinal emittance increases from 2.5 eVs at injection to 8.4 eVs at collision, and the RF voltage is chosen to be 20 MV and the RMS bunch length is 10.07 cm at injection; whereas at collision, the RF voltage is increased to 40 MV to obtain the objective RMS bunch length of 7.55 cm. For the 800-MHz RF at collision, the RMS bunch length is taken as 5.2 cm which has a certain benefit to the luminosity, and the relevant parameters are: the longitudinal emittance 6.4 eVs, RF voltage 52 MV and RMS momentum spread $0.79 \times 10^{-4}$. The corresponding longitudinal impedance threshold is about 1.6 times as that in the case of 400 MHz and the corresponding transverse impedance threshold is about 1.1 times. The IBS growth time for all schemes are acceptable under the condition of the controlled longitudinal emittance growth during acceleration.

**Table 3.** Beam and RF parameters for all RF schemes at SPPC. (Inj: Injection; Col: collision)

| Parameter | Unit | Inj-1 | Inj-2 | Col-1 | Col-2 | Col-3 |
|---|---|---|---|---|---|---|
| Proton energy | [GeV] | 2100 | 2100 | 37500 | 37500 | 37500 |
| Ring circumference | [m] | 100000 | 100000 | 100000 | 100000 | 100000 |
| Bending radius | [m] | 10415.4 | 10415.4 | 10415.4 | 10415.4 | 10415.4 |
| Number of particles per bunch | $10^{11}$ | 1.50 | 1.50 | 1.50 | 1.50 | 1.50 |
| Number of bunches | | 10080 | 10080 | 10080 | 10080 | 10080 |
| Longitudinal emittance ($4\sigma$) | [eVs] | 2.5 | 2.5 | 8.4 | 6.4 | 8.4 |
| Stored energy per beam | [MJ] | 508.7 | 508.7 | 9083.3 | 9083.3 | 9083.3 |
| Slip factor $\eta$ | $10^{-4}$ | 1.014 | 1.014 | 1.016 | 1.016 | 1.016 |
| Revolution frequency | [kHz] | 3.00 | 3.00 | 3.00 | 3.00 | 3.00 |
| RF frequency | [MHz] | 200 | 400 | 400 | 800 | 400+800 |
| Bunch spacing | ns | 25 | 25 | 25 | 25 | 25 |
| Harmonic number | | 66667 | 133333 | 133333 | 266667 | 133333 |
| Total RF voltage | [MV] | 3.7 | 20 | 40 | 52 | 40+20 |
| Synchrotron frequency | [Hz] | 4.1 | 13.6 | 4.55 | 7.33 | 6.43 |

| Bucket area | [eVs] | 5.45 | 4.48 | 26.73 | 10.77 | 20.99 |
| --- | --- | --- | --- | --- | --- | --- |
| Bucket half height ($\Delta E/E$) | [$10^{-3}$] | 0.41 | 0.67 | 0.22 | 0.18 | 0.22 |
| RMS bunch length | [cm] | 18.25 | 10.07 | 7.55 | 5.19 | 6.35 |
| Full bunch length ($4\sigma$) | [ns] | 2.43 | 1.34 | 1.01 | 0.69 | 0.85 |
| Momentum spread $\sigma_\delta$ | $10^{-4}$ | 1.56 | 2.82 | 0.71 | 0.78 | 0.84 |
| Energy loss per turn $U_0$ | [keV] | 0.01 | 0.01 | 1477.88 | 1477.88 | 1477.88 |
| Longitudinal damping time $\tau_\varepsilon$ | [h] | 6681 | 6681 | 1.17 | 1.17 | 1.17 |
| IBS growth time-H | [h] | 70.4 | 55.3 | 166 | 168 | -- |
| IBS growth time-L | [h] | 98.7 | 58.6 | 171 | 118 | -- |

**3.5 Dual harmonic RF systems**

The 800-MHz single-harmonic RF scheme seems to be a good alternative to the SPPC baseline design of 400-MHz RF system, due to a shorter bunch length and higher transverse and longitudinal impedance thresholds. However, the RF bucket occupation by bunch is too full which will have an adverse effect on the bunch storage and collision during physics running of up to 14 h.

Another more promising RF scheme is to use a dual-harmonic RF system combining the basic RF system of 400 MHz and a higher harmonic RF system of 800 MHz. As mentioned in section 3.1, the dual harmonic RF system can work in two different modes: the bunch lengthening mode (BLM) and bunch shortening mode (BSM). Figure 11 presents the synchrotron tune distribution for dual-harmonic RF system with different RF voltage ratios ($k$ value) between 800 MHz and 400 MHz. BLM has relatively smaller synchrotron frequency spread in most cases and zero derivative of synchrotron frequency at certain phase, which is disadvantageous for Landau damping [25]. Furthermore, the average synchrotron tune is smaller, which is unfavorable for controlling the Transverse Mode Coupling Instability [21]. On the other hand, both the synchrotron frequency spread and the average synchrotron tune in BSM are improved compared with the single harmonic RF system at 400 MHz. Besides, the bunch length is slightly shorter, which is helpful to enhance the luminosity. Although the bucket area is relatively smaller, it can be cured by giving a larger voltage on the fundamental harmonic RF if needed. Therefore, the dual-harmonic RF system working in BSM using the combination of two RF systems, one with 400 MHz and 40 MV and the other with 800 MHz and 20 MV, is adopted to mitigate the instabilities and increase the luminosity by producing shorter bunches. Accordingly, a set of corresponding longitudinal dynamics parameters has been designed and optimized, as shown in the last column of table 3 (Col-3). The RMS bunch length reduces to be 6.35 cm from 7.55 cm in the case of 400 MHz alone, and the RMS momentum spread rises to be $0.84 \times 10^{-4}$ from $0.71 \times 10^{-4}$ with the same longitudinal emittance of 8.4 eVs.

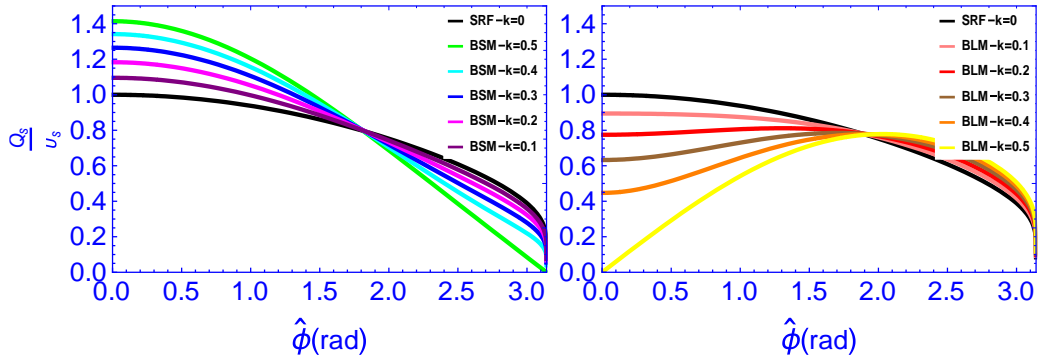

**Figure 11.** Synchrotron tune distribution for dual-harmonic RF system with different RF voltage ratios (*k* value) between 800 MHz and 400 MHz. $\hat{\phi}$ is the maximum amplitude of phase oscillation.

## 4. Parameter design on longitudinal dynamics for the injection chain

The main goal of the injector chain aims at pre-accelerating beams to the injection energy of 2.1 TeV for the SPPC with the required bunch properties such as bunch current, bunch structure, emittance and beam filling period. Therefore, it is crucial to design a set of self-consistent longitudinal dynamics parameters to make the beam stably accelerated at each stage of the injector chain and to achieve the perfect bunch-to-bucket transfer between two neighboring accelerators of the injector chain.

### 4.1 General considerations for the longitudinal dynamics in the injector chain

#### 4.1.1 Injection matching in the longitudinal plane between two accelerators in cascade

The injection and RF capture processes in the longitudinal plane are critical in the different stages of the injector chain, during which major beam loss usually occurs. An almost perfect longitudinal phase-space matching during injection between two neighboring accelerators of the injector chain requires the minimized errors in energy and phase by means of energy and phase correction loops, and the matching in emittance shape and size.

For the beam injection from a proton linac to a synchrotron, in the case of low-intensity beams, the adiabatic RF capture procedure is often practiced to transform the coasting beam formed during the multi-turn injection into a bunched beam. However, for high-intensity beams, the H- stripping and phase space painting method is generally employed to obtain proton accumulation with a large number of injection turns and very low beam loss. As a measure to reduce the longitudinal beam loss, the p-Linac macro pulse is chopped with the p-RCS RF frequency, which enhances the longitudinal phase painting process.

For the beam injection from one synchrotron to the other, the bunch-to-bucket transfer method is utilized, though the two synchrotrons may have different RF frequencies with the latter being a multiple of the former. In this method, the basic premise is that the RF buckets are stationary during the extraction in one synchrotron and the injection in the other synchrotron, to avoid mismatches in beam energy and bunch shape caused by the acceleration buckets. Besides the matches in beam energy and timing or RF phase, the matches in the longitudinal emittance and its shape are also critical. Thus, the RF voltages in the two synchrotrons are bound by the following expression [11, 26]:

$$\frac{V_i}{V_r} = \left(\frac{R_i}{R_r}\right)^2 \frac{h_r|\eta|_i}{h_i|\eta|_r} = \left(\frac{f_{RF,r}}{f_{RF,i}}\right)^2 \frac{h_i|\eta|_i}{h_r|\eta|_r}, \qquad (4.1)$$

where the subscripts *i* and *r* refer respectively to the injection or feeding synchrotron and the receiving synchrotron. One can see that the ratio of the matched RF voltages of two synchrotrons is related to the ratios of RF frequencies, RF harmonic numbers and slippage factors.

### 4.1.2 Space charge effects and its mitigation

The space charge effects are critical in designing and operating high-intensity low-or-medium energy proton synchrotrons, and this is the case in p-RCS and MSS. The space charge effects often drive halo production and emittance growth that will probably lead to important beam losses [27]. The tune shift and spread due to the space charge effects are the key factors in causing beam losses. Tune shift/spread is large when the beam energy is low and will cause the particles to cross dangerous resonances, whereas it will gradually decay with the increase of beam energy. The average tune shift by the direct space charge effect can be described by the well-known Laslett tune-shift formula [28]:

$$\Delta Q = -\frac{r_p n_t}{2\pi \beta^2 \gamma^3 \varepsilon B_f}, \qquad (4.2)$$

where $n_t = h N_b$ is the equivalent total particles in the ring, $\varepsilon$ is the geometric transverse emittance and $B_f = \bar{I}/\hat{I}$ is the bunching factor that is defined by the ratio of the average current $\bar{I}$ to the peak current $\hat{I}$ for a bunched beam. The bunching factor can be derived, for the stationary bucket of dual-harmonic RF system, as [29, 30]:

$$B_f = \frac{2\sin\phi_m - 2\phi_m \cos\phi_m - k/2\,(\sin 2\phi_m - 2\phi_m \cos 2\phi_m)}{2\pi(1 - \cos\phi_m + k/2\,(\cos 2\phi_m - 1))}, \qquad (4.3)$$

where $\phi_m$ is the maximum amplitude of the phase oscillation, and the case of $k=0$ is the bunching factor for the single harmonic RF system.

For the p-RCS and MSS working in the mode to feed beam for the SPPC collision, the normalized RMS transverse emittance is 2.4 πmm·mrad and the bunch population is $1.5\times10^{11}$. The influence of the bunching factor on tune shift due to incoherent space charge at the injection of the p-RCS and MSS is shown in figure 12. One can see that the space charge tune shift in the MSS is much smaller than that in the p-RCS due to much higher injection energy in the former.

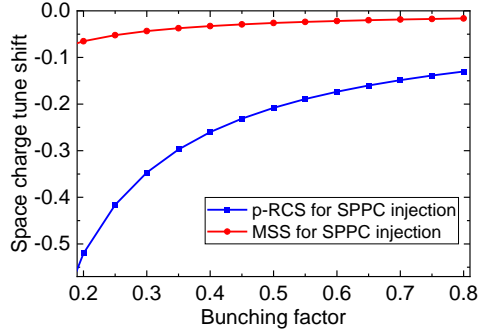

**Figure 12.** Tune shift due to incoherent space charge during the injection of the p-RCS and MSS.

### 4.1.3 Momentum filling factor

A reasonable momentum filling factor (see eq. (3.11)) is also crucial to avoid the serious filamentation and uncontrolled emittance growth of a bunch. However, eq. (3.11) is just an approximation for BLM with $k > 0.5$, since the bucket center will become an unstable fixed point, and two new stable fixed points will occur at $\phi = \pm \arccos(1/2k)$ where the bunch/bucket height just locates [12]. Instead, the momentum filling factor in this case can be derived as:

$$q_p = \sqrt{\frac{k\cos^2\phi_m - \cos\phi_m + 1/4k}{k + 1 + 1/4k}}, k > 0.5. \quad (4.4)$$

Hence, the momentum filling factor for $k > 0.5$ is shown in figure 13 (left), with q and q1 denoting the momentum filling factors obtained from the approximate formula eq. (3.11) and the accurate formula eq. (4.4), respectively. One can see that the difference is negligible for $k = 0.6/0.7$ when the bunch length is much greater than the width of the inner bucket, which is located at the point $q_p = 0$ in figure 13 (left). Therefore, eq. (3.11) can still be regarded as an approximation of the momentum filling factor for $k > 0.5$.

Figure. 13 (right) shows the momentum filling factors for different $k$ values obtained from eq. (3.11). It is clear that for the same bunch length, the momentum filling factor will be smaller for a larger $k$ value, which means safer from beam loss. Nonetheless, the cases $k > 0.5$ have weaker effects on the momentum filling factor for a longer bunch length like $\phi_m > 2$.

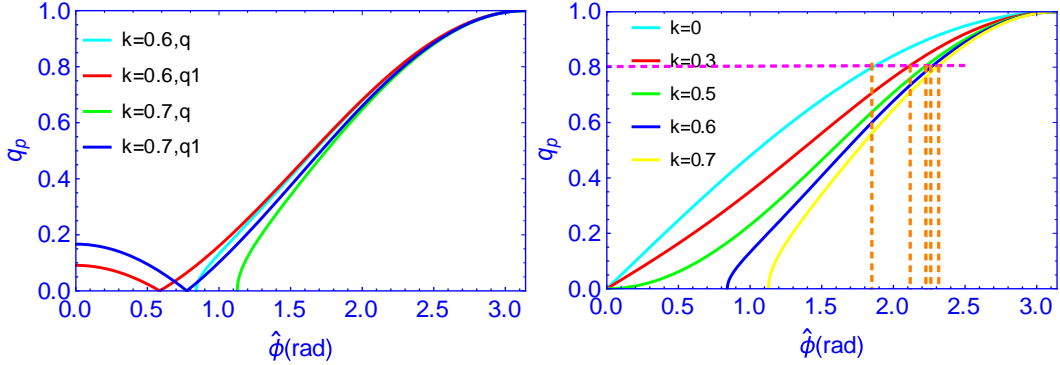

**Figure 13.** The comparison between the accurate formula and approximate formula of the momentum filling factor for the case of $k > 0.5$, with q and q1 denoting the momentum filling factors obtained from the approximate formula eq. (3.11) and the accurate formula eq. (4.4), respectively (left); The momentum filling factor for different $k$ values is obtained from eq. (3.11) (right).

### 4.2 Beam and RF parameters for the injector chain

Taking the above considerations into account, a set of self-consistent longitudinal dynamics parameters for the injector chain was designed, as presented in table 4, and some details are given below. It is worth pointing out that the longitudinal emittance was increased in a controlled way intentionally, with the maximum growth multiple about 3, during each stage of the injector chain, just as that in the LHC complex [31].

#### 4.2.1 SS

The SS ring as the last stage of the injector chain is a special synchrotron with the circumference 7.2 km using superconducting magnets of about 8 Tesla. An RF system of 200 MHz is adopted. The corresponding harmonic number is 4800 and the phase slippage factor is $1.52 \times 10^{-3}$. The bunch at extraction should be matched with that at the SPPC injection. Thus, the longitudinal emittance will be increased in a controlled way from 1.0 eVs at injection to 2.5 eVs at extraction. The RF voltages are 4.5 MV and 4 MV for injection and extraction, repectively, taking into account the momentum filling factor and possible impedance threshold described in section 3.3. The RF power delivered to beam during acceleration is around 2.6 MW. Other longitudinal dynamics parameters were also obtained, as shown in table 4, using the corresponding formulas described in section 3.1.

### 4.2.2 MSS

The general design of MSS must consider the potential for high-power beam applications about several MW in the standalone mode. The MSS lattice with a negative momentum compaction factor of $-7.7\times10^{-4}$ was designed under the restriction of the circumference of about 3.5 km. When serving beams to the SPPC, the instant beam power is 40 kW at injection and 730 kW at extraction. As shown in Fig.12, the space charge tune shift is smaller than 0.1 at the MSS injection, and it will decay with the increase of energy. Thus, the constant 40-MHz single-harmonic RF system during the MSS cycle is used for acceleration and the corresponding harmonic number is 464. The required RF voltage at the MSS extraction is 5.8 MV that is derived from the longitudinal phase-space match with the SS, so are the other corresponding longitudinal dynamics parameters. The RF power delivered to beam during acceleration is about 1.7 MW. Since the momentum filling factor at the MSS extraction is too small, the loss of Landau damping seems extremely severe, with the threshold of the reactive impedance only about 0.2 Ω. A 200-MHz RF system can be added acting as a Landau cavity to address this issue if needed. The longitudinal emittance at the MSS injection is set as 0.5 eVs, thus, the corresponding longitudinal dynamics parameters can be obtained considering the space charge effect with the bunching factor at least 0.2 and the momentum filling factor not above 0.8, as shown in table 4.

### 4.2.3 p-RCS

As the first-stage synchrotron of the SPPC injector chain, the p-RCS is a rapid cycling synchrotron with the repetition of 25 Hz and with the capability of accommodating beam power up to several MW in the standalone mode. When serving beams to the SPPC, the instant beam power is 80 kW at injection and 670 kW at extraction. The space charge tune shift that is most serious at the p-RCS injection can be tolerated not larger than 0.3 due to the relatively low beam power at the injection. Therefore, a dual-harmonic RF system combining the fundamental frequency (36 MHz at injection) RF cavities and the doubled-frequency RF cavities is employed to lengthen the bunch to obtain a bunching factor of larger than 0.4 during the phase of injection and early acceleration, whereas only the fundamental RF is remained at the p-RCS extraction.

The circumference of p-RCS is about 967 m. A special lattice with the negative momentum compaction factor of $-6.75\times10^{-3}$ was designed to obtain the phase slippage factor with a large absolute value, which helps to get a reasonable RF voltage, about 1.46 MV, at the p-RCS extraction. Otherwise, to attain the RF voltage matching during the transfer from the p-RCS to the MSS, the voltage at the p-RCS extraction will be very low, less than one hundred kV that will lead to extremely severe beam loading effect. The initial longitudinal emittance obtained from longitudinal phase space painting of p-Linac beam is set as 0.16 eVs, and the RF voltage is adopted as 1.0 MV, which leads the bunching factor of 0.50 and momentum filling factor of 0.75. With the fast ramping in the sinusoidal waveform, the peak RF power delivered to beam during acceleration is about 1.8 MW. Other parameters related to longitudinal dynamics were also obtained, as presented in table 4.

Table 4. Beam and RF parameters in the SPPC injector chain. (Inj: injection; Ext: Extraction)

| Parameter | Unit | p-RCS | MSS | SS |
|---|---|---|---|---|
| | | Inj / Ext | Inj / Ext | Inj / Ext |
| Proton energy | [GeV] | 1.2 / 10 | 10 / 180 | 180 / 2100 |
| Ring circumference | [m] | 967 | 3478 | 7200 |
| Dipole field | [T] | 0.18 / 1.0 | 0.1 / 1.7 | 0.71 / 8.26 |
| Repetition rate | [Hz] | 25 | 0.5 | 1/30 |
| Bunch intensity | $10^{11}$ | 1.5 | 1.5 | 1.5 |
| Number of bunches | | 112 | 336 | 672 |
| Bunch spacing | [ns] | 27 / 25 | 25 | 25 |
| Longitudinal emittance ($4\sigma$) | [eVs] | 0.16 / 0.5 | 0.5 / 1.0 | 1.0 / 2.5 |
| Revolution frequency | [kHz] | 278.8 / 310.2 | 86.3 / 86.3 | 41.7 |
| Beam power | [MW] | 0.08 / 0.67 | 0.04 / 0.73 | -- |
| Stored beam energy | [MJ] | -- | -- | 2.9 / 33.9 |
| Slip factor $\eta$ | $10^{-3}$ | -199.3 / -14.1 | -8.128 / -0.797 | 1.49 / 1.52 |
| RF frequency | [MHz] | 36+72 / 40 | 40 | 200 |
| Harmonic number | | 128 | 464 | 4800 |
| Total RF voltage | [MV] | 1+0.5 / 1.46 | 3 / 5.8 | 4.5 / 4 |
| Synchrotron frequency | [Hz] | 10163 / 1922 | 1106.7 / 118.5 | 221.93 / 61.96 |
| Bucket area | [eVs] | 0.27 / 2.40 | 2.37 / 42.77 | 5.40 / 17.24 |
| Bucket half height ($\Delta E/E$) | $[10^{-3}]$ | 3.07 / 6.83 | 6.80 / 7.43 | 1.49 / 0.41 |
| Full bunch length ($4\sigma$) | [ns] | 18.78 / 8.25 | 8.25 / 2.75 | 2.75 / 2.43 |
| Momentum spread $\sigma_\delta$ | $10^{-4}$ | 0.96 / 1.76 | 2.46 / 0.64 | 0.64 / 0.16 |
| Momentum filling factor | | 0.75 / 0.49 | 0.50 / 0.17 | 0.76 / 0.69 |
| Bunching factor | | 0.50 / 0.21 | 0.22 / 0.07 | 0.35 / 0.31 |

## 5. Conclusions and discussion

The bunch filling scheme of the SPPC accelerator complex was designed based on various constraints, such as the nominal bunch spacing and the bunch filling factor, the beam-beam effects, and injection/extraction gaps. The final bunch filling factor in the collider rings is 0.756, which is slightly less than 0.8 at LHC but can be enhanced in the future if the shorter rise times of the injection kickers can be applied. A set of self-consistent beam and RF parameters for the SPPC complex were systematically studied and obtained, with the focus of the collider ring on the RF scheme to fulfill the requirements for high luminosity and to mitigate relevant instabilities. The dual-harmonic RF system combining the fundamental harmonic at 400 MHz and a doubled-harmonic at 800 MHz and working on the bunch shortening mode was found beneficial for the SPPC collider rings in mitigating the instabilities and increasing the luminosity by producing shorter bunches at the collision. The longitudinal phase-space matchings between different accelerator stages of the injector chain, the space charge effects and the ratio of bunch to RF bucket and so on, were addressed. A special lattice with the negative momentum filling factor for p-RCS was designed to alleviate the beam loading effect at the extraction and facilitate the longitudinal bunch-to-bucket transfer from p-RCS to MSS. The method of the parameter design of the longitudinal dynamics can also be applicable to other high-energy proton colliders.

There is also other important work remaining to be investigated in the future. First, only single bunch collective instabilities are considered in this paper, other instabilities like longitudinal coupling bunch instability are also of great importance, which need to be specially studied later. Second, the requirement on the reactive impedance at the MSS imposed by the loss of Landau damping seems extremely severe, and should be addressed more carefully. Furthermore, to mitigate the pile-up effects in the detectors and the collective instabilities of single bunch, the possibility of producing a shorter bunch spacing of 5 ns is to be investigated. At last, the longitudinal dynamics for both p-RCS and MSS working in the standalone high-power mode needs a dedicated study.

## Acknowledgments

This work was supported by the National Natural Science Foundation of China (Projects 11575214, 11527811 and 12035017). The authors would like to thank all the members of the SPPC study group for their early contributions to the lattice design and preliminary consideration for each stage of the SPPC complex.